\documentclass[11pt]{article}
\usepackage{amsmath,amssymb,color}
\usepackage{amsfonts}
\newcommand{\hoch}[1]{$\, ^{#1}$}

\newcommand{\auth}{H. L\"u\hoch{\dagger\ddagger} and
Zhao-Long Wang\hoch{\star}}

\def\ft#1#2{{\textstyle{\frac{\scriptstyle #1}{\scriptstyle #2} } }}
\def\fft#1#2{{\frac{#1}{#2}}}
\def\CP{{{\mathbb C}{\mathbb P}}}
\def\0{{\sst{(0)}}}
\def\1{{\sst{(1)}}}
\def\2{{\sst{(2)}}}
\def\3{{\sst{(3)}}}
\def\4{{\sst{(4)}}}
\def\5{{\sst{(5)}}}
\def\6{{\sst{(6)}}}
\def\7{{\sst{(7)}}}
\def\8{{\sst{(8)}}}
\def\sst#1{{\scriptscriptstyle #1}}
\def\oneone{\rlap 1\mkern4mu{\rm l}}

\begin{document}

\begin{flushright}
\hfill{ \
USTC-ICTS-10-01\ \ \ \ }
 %\hfill{
%\bf hep-th/yymmnnn}
\end{flushright}

\vspace{25pt}
\begin{center}
{\large {\bf On M-Theory Embedding of Topologically Massive
Gravity}}

\vspace{15pt}

\auth

\vspace{10pt}

\hoch{\dagger}{\it China Economics and Management Academy\\
Central University of Finance and Economics, Beijing 100081}

\vspace{10pt}

\hoch{\ddagger}{\it Institute for Advanced Study, Shenzhen
University, Nanhai Ave 3688, Shenzhen 518060}

\vspace{10pt}

\hoch{\star}{\it Interdisciplinary Center for Theoretical Study,\\
University of Science and Technology of China, Hefei 230026}

\vspace{40pt}

\underline{ABSTRACT}
\end{center}

We show that topologically massive gravity can be obtained by the
consistent Kaluza-Klein reduction from recently constructed
seven-dimensional gravity with topological terms. The internal
four-manifold should be Einstein with the Pontryagin four-form
constantly proportional to the volume form. We also discuss the
possible lift of the system to $D=11$. This enables us to connect
the mass parameter $\tilde\mu$ in $D=3$ to the M5-brane charge. The
dimensionless quantity $3/(G\tilde \mu)$ is discrete and
proportional to $N$, where $N$ is the number of M5-branes.

\vspace{15pt}

\thispagestyle{empty}

%\pagebreak
%\voffset=0pt
%\setcounter{page}{1}

%\tableofcontents

%\addtocontents{toc}{\protect\setcounter{tocdepth}{2}}

%%%%%%%%%%%%%%%%%%%%%%%%%%%%%%%%%%%%%%%%

\newpage
%%%%%%%%%%%%%%%%%%%%%%%%%%%%%%%%%%%%%%%%

  Although string theory remains to be the most promising candidate
for quantizing gravity, its whole package has been proved to be
difficult to study.  Three-dimensional pure gravity has thus been
constantly attracting people's attention as an intriguing,
tantalizing and much simpler model.  The Einstein-Hilbert action
provides no propagating degree of freedom; however, when a negative
cosmological constant is introduced, the theory admits a non-trivial
BTZ black hole \cite{btz}.  In the 80's topologically massive
gravity (TMG), a higher-derivative theory, was constructed and it
contains one massive particle at the price that the sign of the
Einstein-Hilbert action has to be reversed \cite{Deser:1981wh}. The
sign reversal however would imply that the mass of the BTZ black
hole turns negative.  It was shown that the massive graviton
disappears for certain choice of the cosmological constant and the
mass parameter \cite{stromingeretal}. More three-dimensional
gravities with higher but finite derivatives were recently
constructed and shown to be unitary \cite{Bergshoeff:2009hq,Dalmazi}.

   There may indeed exist self-contained three-dimensional quantum
gravities owing to these unusual properties. It is nevertheless of
great interest to embed them in string theory.  It was demonstrated
that the quantity $c_L-c_R=3/(G\tilde \mu)$ for TMG does not change
under the holographic RG flow \cite{RG}. This result was further
confirmed by the $\beta$-function calculation \cite{ps}. This
strongly suggests that there is a quantization condition for the
dimensionless coupling $G\tilde\mu$. However, it was known that the
homotopy condition associated with the Lorentz Chern-Simons term in
three dimensions is trivial and it gives rise to no quantization
condition for $G\tilde\mu$ \cite{Deser:1981wh}.

    We expect that such a quantization condition has an origin in
higher dimensions.  Indeed, in a seven-dimensional gravity theory
with topological terms, recently constructed in \cite{Lu:2010sj},
the analogous coupling constant is quantized owing to the
non-trivial homotopy condition in $D=7$.

     In this paper, we show that TMG can in fact be obtained by the
consistent Kaluza-Klein reduction from seven-dimensional gravity
with topological terms.  The quantity $G\tilde\mu$ can then be
related to the corresponding one in $D=7$, and hence becomes
quantized. Topological terms in $D=7$ arise naturally in
supergravity coming from the $S^4$ reduction from M-theory,
associated with the anomaly cancelation terms. This may provide a
connection of the mass parameter in $D=3$ to the M5-brane charge,
which is quantized. We shall first demonstrate the embedding of
$D=3$ in $D=7$, and then discuss the embedding in $D=11$.  We shall
finish the paper with a conclusion and further discussions.

    The action for topological gravity in seven dimensions is given by
\cite{Lu:2010sj}.
\begin{equation}\label{d7action}
S=\fft{1}{2\kappa_7^2} \int \Big((R-2\Lambda) {*\oneone} + \mu
\Omega_1^\7 + \nu \Omega_2^\7\Big)\,,
\end{equation}
where
\begin{eqnarray}\label{A1}
\Omega^\7_1 &=&\Omega^{(3)}\wedge {\rm Tr}(\Theta^2)\,,\qquad
\Omega^{(3)}={\rm Tr}(d\Gamma\wedge\Gamma -\ft23\Gamma^3)\,,\cr
\Omega^\7_2 &=&{\rm Tr}(\Theta^3\wedge\Gamma+
 \ft25\Theta^2\wedge\Gamma^3+
 \ft15\Theta\wedge\Gamma^2\wedge\Theta\wedge\Gamma+
 \ft15\Theta\wedge\Gamma^5+\ft{1}{35}\Gamma^7)\,.
\end{eqnarray}
Here, $\Theta$ is the curvature 2-form, defined as $\Theta\equiv
d\Gamma-\Gamma\wedge\Gamma$. In this paper, for a form $X$, the
notation $X^n$ denotes the wedge product of $n$ $X$'s.  The
$\Omega_i^\7$'s are related to the Pontryagin eight-form by
\begin{equation}
P^\8 = \fft{1}{128\pi^4} (d\Omega_1^\7 - 2 d\Omega_2^\7)\,.
\end{equation}

      The equations of motion are given by \cite{Lu:2010sj}
\begin{equation}
 R^{ij}-\ft12g^{ij}R+\Lambda g^{ij}+C_1^{ij}+C_2^{ij}=0\,,
\end{equation}
where
\begin{eqnarray}\label{}
    C_1^{ij}&=&\frac{\delta S_1}{\sqrt{g}\delta
    g_{ij}}=\frac{\mu}{4\sqrt{g}}[\epsilon^{ij_1j_2j_3j_4j_5j_6}
    (R^{i_1}_{~i_2j_1j_2}R_{~i_1j_3j_4}^{i_2} R^{jk}_{~~j_5j_6})_{;k}
    +i\leftrightarrow j]\,,\cr
    C_2^{ij}&=&\frac{\delta S_2}{\sqrt{g}\delta
    g_{ij}}=\frac{\nu}{4\sqrt{g}}[\epsilon^{ij_1j_2j_3j_4j_5j_6}
    (R^{k}_{~i_1j_1j_2}R^{i_1}_{~i_2j_3j_4}R^{ji_2}_{~~j_5j_6})_{;k}
    +i\leftrightarrow j]\,.
\end{eqnarray}

We now perform the Kaluza-Klein dimensional reduction on an internal
four-space with fixed breathing mode. We first consider the case
with positive cosmological constant and let $2\Lambda=30\lambda^2$.
The metric ansatz is
\begin{equation}\label{redansatz}
ds_7^2 = ds_3^2 + \fft{1}{\lambda^2} ds_{4}^2\,.
\end{equation}
We find that the consistency of the equations of motion requires
that $ds_4^2$ is an Einstein space with $R_{ij}=6 g_{ij}$.
Furthermore, the Pontryagin four-form in $ds_4^2$ satisfies
\begin{equation}\label{strongcon}
P^\4\equiv \fft{1}{8\pi^2} {\rm Tr}(\Theta\wedge \Theta)= c\, {\cal
V}_\4\,,
\end{equation}
where ${\cal V}_\4$ is the volume form for $ds_4^2$ and $c$ {\it
has} to be a constant. It is worth emphasizing that these conditions
on the internal four-space can be easily missed if one merely
substitutes the ansatz into the action.  The condition
(\ref{strongcon}) is particularly strong, but it can be satisfied.
For example, homogeneous spaces clearly satisfy (\ref{strongcon}).

      Let $\eta$ denote the Pontryagin instanton number, defined by
\begin{equation}
\eta=\int P^\4\,.
\end{equation}
We find that the effective action in three dimensions is given by
%%%%
\begin{equation}\label{d3action}
S=\fft{1}{2\kappa_3^2} \int \Big( (R - 6 \lambda^2){*\oneone} +
\fft{1}{2\tilde \mu} \Omega^\3\Big)\,,
\end{equation}
where
\begin{equation}
\fft{1}{\tilde\mu} = \fft{{32}\pi^2\lambda^4\eta\mu}{V_4}\,.
\end{equation}
Here $V_4$ is the volume of $ds_4^2$ and $\kappa_3^2=
\lambda^4\kappa_7^2/V_4$. Note that the three-dimensional Newton's
constant is given by $2\kappa_3^2=16\pi G$. We would like to
emphasize again that only one-half of the coupling $1/\tilde\mu$ can
be obtained if one merely substitutes the ansatz into the
seven-dimensional action. Now we arrive at the action of
topologically massive gravity in $D=3$. The (A)dS$_3$ length is
given by
\begin{equation}
\fft{1}{\ell} = \sqrt3 \lambda\,.
\end{equation}
The coupling constant of the topological term depends on the
seven-dimensional coupling constant and the topology of the internal
spaces.  For the round $S^4$, we have $\eta=0$ and hence the
topological term vanishes.  As another example, let us consider the
``round'' $\CP^2$ with the desired cosmological constant.  The
metric is given by
\begin{equation}\label{cp2met}
ds^2 = d\xi^2 + \ft14 \sin^2\xi \cos^2\xi (d\psi + \cos\theta\,
d\phi)^2 + \ft14 \sin^2\xi (d\theta^2 + \sin^2\theta\, d\phi^2)\,.
\end{equation}
Thus, we have $V_4 = \ft12\pi^2$ and $\eta = 3$. The corresponding
$\tilde \mu$ is given by
\begin{equation}\label{cp2mu}
\fft{1}{\tilde \mu} = 192\mu\lambda^4\,.
\end{equation}

  For the case with negative cosmological constant $2\Lambda =-30
\lambda^2$, the internal space is Einstein with $R_{ij}=-6g_{ij}$,
together with the condition (\ref{strongcon}). If we allow the
internal space to be non-compact, such an example can be easily
found. We can use the non-compact anti-$\CP^2$ metric, obtained by
letting $\xi\rightarrow {\rm i}\, \xi$ and $ds^2\rightarrow -ds^2$
in (\ref{cp2met}). The resulting action is then given by
(\ref{d3action}) but with $6\lambda^2\rightarrow -6\lambda^2$. The
parameter $\mu$ is still given by (\ref{cp2mu}).

    If we insist that internal manifolds must be compact, the negative
cosmological constant makes it impossible to find an explicit metric
to satisfy (\ref{strongcon}), since there can be no globally
homogenous compact manifold with negative cosmological constant.
However, equations of motion deal with only local properties, and a
manifold can be locally homogeneous even though it is not globally.
An example of this type is the two-dimensional Riemann surfaces with
high genuses. There is no reason why a similar phenomenon could not
arise in four dimensions.

   If the cosmological constant in $D=7$ vanishes, the internal space
$ds_4^2$ is then Ricci-flat. Since there can be no non-trivial
Ricci-flat homogeneous space even locally, the lower-dimensional
theory cannot inherit any topological term.  Thus the cosmological
constant appears to be crucial for topologically massive gravity to
be embedded consistently in $D=7$.

   In a generic bosonic theory, the coupling constants
$\mu, \nu$ in $D=7$ are unrelated; $\mu$ is continuous whilst $\nu$
is quantized by the homotopy condition \cite{Lu:2010sj}.  Since only
the $\mu$ term would yield the topological term in $D=3$, it follows
that there is no quantization condition for the mass parameter in
$D=3$.  However, for theories where $\mu$ and $\nu$ are related, the
quantization condition in $D=7$ then provides a quantization
condition for $D=3$.  In particular, as was pointed out in
\cite{Lu:2010sj}, topological terms in seven dimensions arise
naturally in supergravity from the $S^4$ reduction of M-theory. In
this case $\mu$ and $\nu$ are related to form the anomaly polynomial
for M5-brane \cite{Duff,Dixon,Ginsparg}. The corresponding mass
parameter in $D=3$ is then quantized.

To see this in detail, let us write the relevant terms in the $D=11$
supergravity action.
\begin{equation}
S=\fft{1}{2\kappa_{11}^2}\int R{*\oneone} - \ft12 {*F_\4}\wedge F_\4
- \ft16 A_\3\wedge(F_\4\wedge F_\4 + 12\kappa_{11}^2 T_3 \, X_\8) +
{\rm more}\,,
\end{equation}
where
\begin{equation}
X_\8 = \fft{1}{192 (2\pi)^4} (\ft14 d\Omega_1^\7 - d\Omega_2^\7)\,,
\end{equation}
and $T_3$ is the M2-brane tension, given by
\begin{equation}
T_3=\fft{1}{(2\pi)^2 \ell_p^3}\,.
\end{equation}
We have introduced the Planck length $\ell_p$, related to
$\kappa_{11}$ by $2\kappa_{11}^2=(2\pi)^8\, \ell_p^9$.

     Eleven-dimensional supergravity admits an AdS$_7\times S^4$
vacuum solution, given by
\begin{eqnarray}
ds_{11}^2 &=& ds_{AdS_7}^2 + Q^{2/3} d\Omega_4^2\,,\cr
F_\4&=&3Qd\Omega_\4\,.
\end{eqnarray}
The AdS$_7$ radius is $2Q^{1/3}$. In other words, we have
$\lambda=\ft12 Q^{-1/3}$.  Here the constant $Q$ is related to the
M5-brane charge, which is quantized as $Q=N\pi\ell_p^3$, where $N$
is the number of M5-branes.

     Neglecting the high-order correction terms, the reduction on
the $S^4$ with fixed breathing mode gives rise to $D=7$ pure gravity
with a cosmological constant, namely
\begin{equation}
S=\fft{1}{2\kappa_7^2} \int (R + 30 \lambda^2)  {*\oneone}\,,
\end{equation}
where
\begin{equation}
\kappa_7^2 = \fft{\kappa_{11}^2}{Q^{4/3} \omega_4}\,.
\end{equation}
Here $\omega_4=\ft83\pi^2$ is the volume of a unit $S^4$. Note that
the theory has a negative cosmological constant. We now consider the
reduction including the $X_\8$ as well. The resulting action is that
of seven-dimensional topological gravity (\ref{d7action}) with
\begin{equation}
\mu=\fft{(2\pi)^2 \ell_p^5}{2^8 (\pi N)^{1/3}}\,,\qquad \nu=-4\mu\,.
\end{equation}
The homotopy condition associated with the $\nu$ term, discussed in
\cite{Lu:2010sj}, is given by
\begin{equation}
\fft{64\pi^4 \nu}{2\kappa_7^2} =\fft{\pi N}{24}=2\pi\, n\,,
\end{equation}
for integer $n$.  Thus we have
\begin{equation}
N=48n\,.
\end{equation}

      Utilizing the previous reduction from $D=7$ to $D=3$, we can
now embed $D=3$ topological gravity in M-theory. The parameters in
three dimensions can be derived from the M-theory parameters, given
by
\begin{equation}
2\kappa^2_3=\fft{6\,\pi^6\ell_p}{(\pi
N)^{8/3}V_4}\,,\qquad\ell=\fft{2}{\sqrt3}(\pi
N)^{\fft13}\ell_p\,,\qquad \fft{1}{\tilde\mu} =\fft{\pi^4\eta
\ell_p}{2^5 (\pi N)^{5/3} V_4}\,.\label{m5mass}
\end{equation}
The central charges of the boundary conformal field theory were
computed in \cite{central1,central2}, and they can now be expressed
in terms of M-theory parameters:
\begin{eqnarray}\label{centralcharge}
c_L&=&\fft{12\,\pi\ell}{\kappa_3^2} \left(1-\fft1{\tilde \mu\,
\ell}\right) =\fft{8V_4}{\sqrt3\,\pi^2}
N^3\left(1-\fft{\pi^2\eta}{2^6V_4}\fft1{N^2}\right)\,, \cr
c_R&=&\fft{12\,\pi\ell}{\kappa_3^2}\left(1+\fft1{\tilde \mu\,
\ell}\right) =\fft{8V_4}{\sqrt3\,\pi^2}N^3\left(1+
\fft{\pi^2\eta}{2^6V_4}\fft1{N^2}\right)\,.
\end{eqnarray}
An interesting observation is that the quantity $c_L-c_R=-\eta
N/(4\sqrt3)$ depends only on the quantized charge $N$ and the
topological charge $\eta$. Therefore, it is invariant under any
continuous perturbation, which is consistent with the fact \cite{RG}
that $c_L-c_R$ does not change under the holographic RG flow.  On
the other hand, the quantity $c_L+c_R\sim N^3 V_4$ suggests that the
breathing scalar mode may be turned on by the RG flow.

   We now examine the consistency and the validity of our reduction.
There are more higher-order correction terms in $D=11$. It is clear
that the type of $A{\rm Tr} (F^2){\rm Tr}(\Theta^2)$ gives no
contribution to the equations of motion with our reduction ansatz.
If all the $R^4$ terms were known in $D=11$, we would expect that
our dimensional reduction procedure would give rise to
higher-derivative supergravity in three dimensions. It would be of
great interest to investigate whether there exists a limit for which
the Lorentz Chern-Simons term dominates. The explicit demonstration
can be difficult because not all the $R^4$ terms have been known.
The fact that TMG can be supersymmetrized in its own right with the
minimum supersymmetry suggests that such a truncation to TMG may
exist.

    To summarize, we have demonstrated that topologically massive
gravity in $D=3$ can be obtained from topological gravity in $D=7$
by the consistent Kaluza-Klein reduction on an internal Einstein
manifold with the Pontryagin four-form satisfying (\ref{strongcon}).
This procedure can be easily generalize to establish a recursive
reduction relation. Topological gravities exist in $4k+3$
dimensions, with $(k+1)$ topological terms. The $k$ terms can be
related to those in $4k-1$ dimensions by
$\Omega^{(4k+3)}=\Omega^{(4k-1)}\wedge {\rm Tr}(\Theta^2)$, and the
extra one is given by $d\Omega^{(4k+3)}={\rm Tr} (\Theta^{2(k+1)})$.
Thus all but the extra one can be reduced and give rise to all the
lower-dimensional topological terms.

   A physically more relevant observation is that topologically
massive gravity can also be embedded in M-theory, with the mass
parameter $\tilde \mu$ now determined by the number of M5-branes,
which has to be large for the truncation approximation to be valid.
The condition for chiral topological gravity, namely $\tilde \mu\,
\ell = 1$ \cite{stromingeretal}, now becomes
\begin{equation}
\fft{\eta}{V_4} = \fft{2^6 N^2}{\sqrt3 \pi^2}\,,
\end{equation}
where $\eta$ is the Pontryagin instanton number and $V_4$ is the
volume of the internal Einstein four-manifold that satisfies $R_{ij}
= -6 g_{ij}$ and (\ref{strongcon}).  It is of great interest to
investigate whether such a manifold could arise mathematically. The
embedding suggests that the dual conformal field theory of
three-dimensional topological gravity is the six-dimensional
conformal field theory of M5-brane wrapped on the Einstein
four-manifold. This proposal is consistent with the leading behavior
of the central charges in (\ref{centralcharge}) since the degree of
freedom of M5 branes is expected to proportion with $N^3$ for large
$N$.

\section*{Acknowledgement}

We are grateful to Nikolay Bobev, Gary Gibbons, Yi Pang, Chris Pope,
and Ergin Sezgin for useful discussions. Z.L.W.~is supported in part
by grants from the Chinese Academy of Sciences, a grant from 973
Program with grant No: 2007CB815401 and grants from the NSF of China
with Grant No: 10588503 and 10535060.

%%%%%%%%%%%%%%%%%%%%%%%%%%%%%%%%%%%%%%%%%%%%%%%%%%%%%%%%%%%%%%%%%%%%%%%

\end{document}